\documentclass[prd,preprint,superscriptaddress,amsmath,amssymb]{revtex4}

\usepackage{graphicx,color,slashed}
\newcommand{\nc}{\newcommand}

\nc{\eps}{\varepsilon}
\nc{\vp}{\varphi}
\nc{\tvp}{\widetilde{\varphi}}
\nc{\D}{\mbox{$\not\!\!D$}}
\nc{\Db}{\mbox{${\raisebox{2mm}{\boldmath ${}^\leftarrow$}\hspace{-4mm} D}$}}
\nc{\Dfb}{\mbox{$\raisebox{2mm}{\boldmath ${}^\leftrightarrow$}\hspace{-4mm} D$}}
\nc{\vpj }{\mbox{${\vp^\dag i\,\raisebox{2mm}{\boldmath ${}^\leftrightarrow$}\hspace{-4mm} D_\mu\,\vp}$}}
\nc{\vpjt}{\mbox{${\vp^\dag i\,\raisebox{2mm}{\boldmath ${}^\leftrightarrow$}\hspace{-4mm} D_\mu^{\,I}\,\vp}$}}

\begin{document}

{\begin{flushright}{KIAS-P18113}
\end{flushright}}

\title{\bf Top-quark flavor-changing $tqZ$ couplings and rare $\Delta F=1$ processes}

\author{Chuan-Hung Chen}
\email{physchen@mail.ncku.edu.tw}
\affiliation{Department of Physics, National Cheng-Kung University, Tainan 70101, Taiwan}

\author{Takaaki Nomura}
\email{nomura@kias.re.kr}
\affiliation{School of Physics, KIAS, Seoul 02455, Korea}

\date{\today}

\begin{abstract}
We  study the impacts of  anomalous $tqZ$ couplings ($q=u,c$), which lead to the $t\to q Z$ decays, on  low energy flavor physics. It is found that the $tuZ$-coupling effect can significantly affect the rare $K$ and $B$ decays, whereas the $tcZ$-coupling effect is small.  Using the ATLAS's branching ratio (BR) upper bound of $BR(t\to uZ) < 1.7\times 10^{-4}$,   the influence of the anomalous $tuZ$-coupling on the rare decays can be found as follows: (a) The contribution to the Kaon direct CP violation  can be up to $Re(\epsilon'/\epsilon) \lesssim 0.8 \times 10^{-3}$; (b) $BR(K^+\to \pi^+ \nu \bar \nu) \lesssim 12 \times 10^{-11}$ and $BR(K_L \to \pi^0 \nu\bar \nu)\lesssim 7.9 \times 10^{-11}$; (c) the BR for $K_S \to \mu^+ \mu^-$ including the long-distance effect can be enhanced by $11\%$ with respect to the standard model result, and (d) $BR(B_d \to \mu^+ \mu^-) \lesssim 1.97 \times 10^{-10}$. In addition, although $Re(\epsilon'/\epsilon)$ cannot be synchronously  enhanced with  $BR(K_L\to \pi^0 \nu \bar\nu)$ and $BR(K_S\to \mu^+ \mu^-)$ in the same region of the CP-violating phase,  the values of $Re(\epsilon'/\epsilon)$, $BR(K^+ \to \pi^+ \nu \bar\nu)$, and $BR(B_d \to \mu^+ \mu^-)$ can be simultaneously increased. 

\end{abstract}

\maketitle

\section{Introduction}

Top-quark flavor changing neutral currents (FCNCs) are extremely suppressed  in the standard model (SM) due to the Glashow-Iliopoulos-Maiani (GIM)  mechanism~\cite{Glashow:1970gm}. The branching ratios (BRs) for the $t\to q (g, \gamma, Z,h)$ decays with $q=u, c$ in the SM are of order  of $10^{-12}-10^{-17}$~\cite{AguilarSaavedra:2004wm,Abbas:2015cua}, and these results are far below the detection limits of LHC,  where the expected sensitivity in the high luminosity (HL)  LHC  for an integrated luminosity of 3000 fb$^{-1}$ at $\sqrt{s}=14$ TeV is  in the range $10^{-5}-10^{-4}$~\cite{ATLAS:2013hta,ATLAS:2016}. 
Thus, the top-quark flavor-changing processes can  serve as good candidates for investigating  the new physics effects.  Extensions of the SM, which can reach the HL-LHC sensitivity, can be found in~\cite{Abraham:2000kx,Eilam:2001dh,AguilarSaavedra:2002kr,Agashe:2006wa,Fox:2007in,Li:2011fza,Li:2011af,Gong:2013sh,Yuan:2010vk,tqZ,Gaitan:2017tka,Mandrik:2018gud}.

Using the data collected with an integrated luminosity of 36.1 fb$^{-1}$ at $\sqrt{s}=13$ TeV, ATLAS reported the current strictest upper limits on the BRs for $t\to q Z$ as~\cite{Aaboud:2018nyl}:
 \begin{align}
 BR(t\to u Z)  & < 1.7 \times 10^{-4} \,, \nonumber \\
 BR(t\to c Z) & < 2.4 \times 10^{-4}\,.
 \end{align}
Based on the current upper bounds, we model-independently study the implications of anomalous $tqZ$ couplings in the low energy flavor physics. It is found that the $tqZ$ couplings through the $Z$-penguin diagram can significantly affect the rare decays in $K$ and $B$ systems, such as $\epsilon'/\epsilon$, $K\to \pi \nu \bar\nu$, $K_S \to \mu^+ \mu^-$, and $B_d \to \mu^+ \mu^-$. Since the gluon and photon in the top-FCNC decays are on-shell,  the contributions from the  dipole-operator transition currents are small. In this study we thus focus on the $t\to qZ$ decays, especially the $t\to u Z$ decay.

From a phenomenological perspective,  the importance of investigating the influence of these rare decays are stated as follows: 
The inconsistency in $\epsilon'/\epsilon$  between theoretical calculations and experimental data was recently found based on two analyses: (i) The RBC-UKQCD collaboration obtained the lattice QCD result  with~\cite{Blum:2015ywa,Bai:2015nea}:
 \begin{equation}
 Re(\epsilon'/\epsilon) = 1.38(5.15)(4.59) \times 10^{-4}\,,
 \end{equation}
where the numbers in brackets denote the errors. (ii) Using a large $N_c$ dual QCD~\cite{Buras:1985yx,Bardeen:1986vp,Bardeen:1986uz,Bardeen:1986vz,Bardeen:1987vg}, the authors in~\cite{Buras:2015xba, Buras:2015yba}  obtained:
 \begin{equation}
 Re(\epsilon'/\epsilon)_{\rm SM} =  (1.9 \pm 4.5)\times 10^{-4}\,.
 %
 \end{equation}
 Note that the authors in~\cite{Gisbert:2017vvj}  could obtain $Re(\epsilon'/\epsilon) = (15 \pm 7) \times 10^{-4}$ when the short-distance (SD) and long-distance (LD) effects are considered. 
Both RBC-UKQCD and DQCD results show that the theoretical calculations exhibit  an over $2\sigma$ deviation from the experimental data of $Re(\epsilon'/\epsilon)_{\rm exp}=(16.6 \pm 2.3)\times 10^{-4}$, measured by NA48~\cite{Batley:2002gn} and KTeV~\cite{AlaviHarati:2002ye,Abouzaid:2010ny}.   Based on the results, various extensions of the SM proposed to resolve the anomaly can be found in~\cite{Buras:2015qea,Buras:2015yca,Buras:2015kwd,Buras:2015jaq,Tanimoto:2016yfy,Buras:2016dxz,Kitahara:2016otd,Endo:2016aws,Bobeth:2016llm,Cirigliano:2016yhc,Endo:2016tnu,Bobeth:2017xry,Crivellin:2017gks,Bobeth:2017ecx,Haba:2018byj,Buras:2018lgu,Chen:2018ytc,Chen:2018vog,Matsuzaki:2018jui,Haba:2018rzf,Aebischer:2018rrz,Aebischer:2018quc,Aebischer:2018csl,Chen:2018dfc,Chen:2018stt}. 
We find that  the direct Kaon CP violation  arisen from the $tuZ$-coupling can be $\epsilon'/\epsilon\lesssim 0.8 \times 10^{-3}$  when  the bound of  $BR(t\to uZ)< 1.7 \times 10^{-4}$ is satisfied.

Unlike $\epsilon'/\epsilon$, which strongly depends on the hadronic matrix elements,  the calculations of $K^+ \to \pi^+ \nu \bar\nu$ and $K_L \to \pi^0 \nu \bar\nu$ are theoretically clean and  the SM results can be found as~\cite{Bobeth:2016llm}:
 \begin{align}
 BR(K^+\to \pi^+ \nu \bar \nu) &= (8.5 ^{+1.0}_{-1.2}) \times 10^{-11} \,, \\
 BR(K_L \to \pi^0 \nu\bar \nu) & = (3.2^{+1.1}_{-0.7})\times 10^{-11}\,,
 \end{align}
  where the QCD corrections at the next-to-leading-order (NLO)~\cite{Buchalla:1993bv,Misiak:1999yg,Buchalla:1998ba} and NNLO~\cite{Gorbahn:2004my,Buras:2005gr,Buras:2006gb} and the electroweak corrections at the NLO~\cite{Buchalla:1997kz,Brod:2008ss,Brod:2010hi} have been calculated. In addition to their sensitivity to new physics, $K_L \to \pi^0 \nu \bar\nu$ is a CP-violating process and its BR indicates  the CP-violation effect. The current experimental situations are $BR(K^+\to \pi^+ \nu \bar \nu)^{\rm exp} =(17.3^{+11.5}_{-10.5})\times 10^{-11}$~\cite{Artamonov:2008qb} and $BR(K_L \to \pi^0 \nu\bar \nu)^{\rm exp} < 2.6\times 10^{-8}$~\cite{Ahn:2009gb}.  
 The NA62 experiment at CERN is intended  to measure the BR for $K^+ \to \pi^+ \nu \bar\nu$ with  a $10\%$ precision~\cite{Rinella:2014wfa,Moulson:2016dul},  and the KOTO experiment at J-PARC  will observe the $K_L \to \pi^0 \nu \bar\nu$ decay~\cite{Komatsubara:2012pn,Beckford:2017gsf}. In addition, the KLEVER experiment at CERN starting in Run-4 could observe the BR  of $K_L \to \pi^0 \nu \bar \nu$ to $20\%$ precision~\cite{Moulson:2018mlx}.  
 Recently, NA62 reported its first result using the 2016 taken data and  found that  one candidate event of $K^+\to \pi^+ \nu \bar \nu$ could be observed, where the corresponding BR upper bound  is given  by $BR(K^+\to \pi^+ \nu \bar\nu)< 14 \times 10^{-10}$ at a $95\%$ confidence level (CL)~\cite{Velghe:2018erh}. We will show that the anomalous $tuZ$-coupling can lead to  $BR(K^+\to \pi^+ \nu \bar \nu) \lesssim 12 \times 10^{-11}$ and $BR(K_L \to \pi^0 \nu\bar \nu)\lesssim 7.9 \times 10^{-11}$.  It can be seen that NA62, KOTO, and KLEVER experiments can further constrain the $tuZ$-coupling using the designed sensitivities.

Another important CP violating process is $K_S \to \mu^+\mu^-$, where the SM prediction including the long-distance (LD) and short-distance (SD) effects is given as $BR(K_S \to \mu^+ \mu^-)=(5.2 \pm 1.5) \times 10^{-12}$~\cite{Ecker:1991ru,Isidori:2003ts,DAmbrosio:2017klp}. The current upper limit from LHCb is $BR(K_S\to \mu^+ \mu^-) <  0.8 (1.0) \times 10^{-9}$ at a 90\%(95\%) CL.  It is expected that using the LHC Run-2 data, the LHCb sensitivity can be improved to  $[4,\, 200]\times 10^{-12}$ with 23 fb$^{-1}$ and to $[1,\, 100]\times 10^{-12}$ with 100 fb$^{-1}$~\cite{Dettori:UKF2017}. Although the $tuZ$-coupling can significantly enhance the SD contribution of $K_S\to \mu^+ \mu^-$, due to LD dominance,  the increase of $BR(K_S \to \mu^+ \mu^-)_{\rm LD + SD}$ can be up to $11\%$.

It has been found that the $tuZ$-coupling-induced  $Z$-penguin can significantly enhance the  $B_d\to \mu^+ \mu^-$ decay, where the SM prediction  is given by $BR(B_d \to \mu^+ \mu^-)=(1.06\pm 0.09)\times 10^{-10}$~\cite{Bobeth:2013uxa}. From the data, which combine the full Run I data with the results of  26.3 fb$^{-1}$ at $\sqrt{s}=13$ TeV,   ATLAS reported the upper limit  as $BR(B_d \to \mu^+ \mu^-) < 2.1 \times 10^{-10}$~\cite{Aaboud:2018mst}. In addition, the result combined CMS and LHCb was reported as $BR(B_d \to \mu^+ \mu^-) =(3.9^{+1.6}_{-1.4}) \times 10^{-10}$~\cite{CMS:2014xfa}, and LHCb recently obtained the upper limit of $BR(B_d \to \mu^+ \mu^-) < 3.4 \times 10^{-10}$~\cite{Aaij:2017vad}. 
 It can be seen that the measured sensitivity is close to the SM result. We find that using the current upper limit of $BR(t\to u Z)$,  the $BR(B_d \to \mu^+ \mu^-)$ can be enhanced up to  $1.97 \times 10^{-10}$, which is close to the ATLAS upper bound. 

The paper is organized as follows: In Sec.~II, we introduce the effective interactions for $t\to qZ$ and derive the relationship between the $tqZ$-coupling and $BR(t\to qZ)$. The  $Z$-penguin FCNC processes induced via the anomalous $tqZ$ couplings are given  in Sec.~III. The influence on $\epsilon'/\epsilon$ is shown in the same section. The $tqZ$-coupling contribution to the other  rare $K$ and $B$ decays is shown in Sec.~IV. A summary is given in Sec.~V.

\section{  Anomalous $tqZ$ couplings and their constraints}

Based on the prescription in~\cite{AguilarSaavedra:2004wm},  we write the anomalous $tqZ$ interactions as:
\begin{align}
-{\cal L}_{t q Z} & = \frac{g}{2 c_W} \bar q  \gamma^\mu \left( \zeta^L_q P_L + \zeta^R_q P_R \right) t Z_\mu + H.c. \,,
 \label{eq:L_tqZ}
\end{align}
where $g$ is the $SU(2)_L$ gauge coupling; $c_W=\cos\theta_W$ and $\theta_W$ is the Weinberg angle; $P_{L(R)}=(1\mp \gamma_5)/2$, and  $\zeta^{L(R)}_q$ 
denote the dimensionless effective couplings and represent the new physics effects. In this study, we mainly  concentrate the impacts of the $tqZ$ couplings on the low energy flavor physics, in which the rare  $K$ and $B$ decays are induced through the penguin diagram. 
 The rare $D$-meson processes, such as $D-\bar D$ mixing and $D\to \ell \bar\ell$,  can be induced through the box diagrams; however, the processes in $D$ system can always be suppressed by taking one of the involved anomalous couplings, e.g. $tcZ$, to be small.   Thus, in the following analysis, we focus on the study in the rare $K$ and $B$ decays. In order to  study the influence on the Kaon  CP violation, we take $\zeta^{L, R}_q $ as complex parameters, and the new CP violating phases are defined as $\zeta^{\chi}_q = |\zeta^{\chi}_q| e^{-i\theta^\chi_q}$ with $\chi=L,R$. 
 The top anomalous couplings in Eq.~(\ref{eq:L_tqZ})  can basically arise
 from the dimension-six operators in the SM effective field theory (EFT), where the theory with new physics effects obeys  the  $SU(2)_L\times U(1)_Y$ gauge symmetry. For clarity, we show the detailed analysis for the left-handed quark couplings in Appendix. It can be found that the couplings in  Eq.~(\ref{eq:L_tqZ}), which are generated from the SM-EFT, are not completely excluded by the low-energy flavor physics when the most general couplings are applied. The case with the strict constraints can be found in~\cite{Fox:2007in}.  In addition to the SM-EFT~\cite{Buchmuller:1985jz,AguilarSaavedra:2009mx,Grzadkowski:2010es},  the  top anomalous $tqZ$ couplings can be induced from the lower dimensional operators in the extension of the SM, such as $SU(2)$ singlet vector-like up-type quark model~\cite{AguilarSaavedra:2002kr}, extra dimensions~\cite{Agashe:2006wa}, and  generic two-Higgs-doublet model~\cite{Gaitan:2017tka}.  Hence, in this study,  we  take $\zeta^\chi_q$ are the free parameters and  investigate the implications of the sizable $\zeta^\chi_q$ effects without exploring their producing mechanism.  
%


Using the interactions in Eq.~(\ref{eq:L_tqZ}), we can calculate the BR for $t\to q Z$ decay. Since our purpose is to examine whether the anomalous $tqZ$-coupling can give sizable contributions to the rare $K$ and $B$ decays when the current upper bound of $BR(t\to q Z)$ is satisfied, we express  the parameters $\zeta^{L,R}_q$ as a function of $BR(t \to q Z)$  to be:
\begin{align}
\sqrt{ \left|\zeta^L_q \right|^2 + \left|\zeta^R_q \right|^2} & = \left(  \frac{BR(t\to q Z)}{C_{tqZ} }\right)^{1/2} \,, \nonumber \\
C_{tqZ} & =   \frac{G_F m^3_t}{16 \sqrt{2} \pi \Gamma_t} \left( 1- \frac{m^2_Z}{m^2_t}\right)^2\left( 1+ 2\frac{m^2_Z}{m^2_t} \right) \,.
\end{align}
For the numerical analysis,  the relevant input values are shown in Table~\ref{tab:inputs}. 
Using the  numerical inputs, we obtain $C_{tqZ} \approx 0.40$. When $BR(t\to u(c) Z) < 1.7 (2.3) \times 10^{-4}$ measured by ATLAS are applied, the upper limits on $\sqrt{|\zeta^L_{u(c)}|^2 + |\zeta^{R}_{u(c)}|^2}$ can be respectively obtained as:
 \begin{align}
 \sqrt{|\zeta^L_{u}|^2 + |\zeta^{R}_{u}|^2} < 0.019\,, \nonumber \\
 \sqrt{|\zeta^L_{c}|^2 + |\zeta^{R}_{c}|^2} < 0.022\,. \label{eq:zeta_upper_limit}
 \end{align}
Since the current measured results of the $t\to (u, c) Z$ decays are close each other, the bounds on $\zeta^\chi_{u}$ and $\zeta^{\chi}_c$ are very similar. We note that BR cannot determine the CP phase; therefore, $\theta^{\chi}_{u}$ and $\theta^{\chi}_c$ are  free parameters. 

\begin{table}[htp]
\caption{Inputs for the numerical estimates. }
\begin{tabular}{cccc} \hline \hline
$m_s=0.109$ GeV & $m_d=5.10$ MeV& $m_c=1.3$ GeV & $m_t(m_t) =165$ GeV    \\ \hline 
$m^{\rm pole}_t=172$ GeV & $m_W=80.38$ GeV &  $\Gamma_t =1.43$ GeV &  $m_K=0.498$ GeV   \\ \hline 
 $m_{B_{d}}=5.28$ GeV & $V_{ud,tb,cs}\approx 1$ & $V_{td}=0.0088 e^{-i 23^{\circ}}$ &  $V_{ts}=-0.041$  \\ \hline 
  $V_{us}=0.225$ & $V_{cd}=-0.225$ & $\sin^2\theta_W=0.23$ &  $f_\pi=0.13$ GeV \\ \hline 
 $f_K=0.16$ GeV & $f_{B}=0.191$ GeV &  $|\epsilon_K|=2.228\times 10^{-3}$ &   $\tau_{K_S(B)}=89.5(1.52)\times 10^{-12}$ s \\ \hline\hline
\end{tabular}
\label{tab:inputs}
\end{table}%

\section{ Anomalous $tqZ$ effects on $\epsilon'/\epsilon$ }

\begin{figure}[phtb]
\includegraphics[scale=0.7]{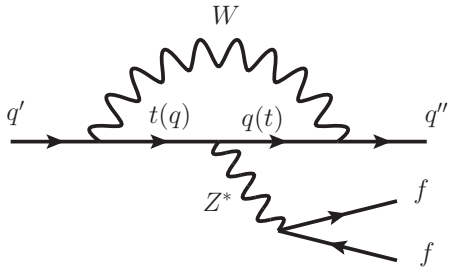}
 \caption{ Sketched Feynman diagram for $q'\to q'' f \bar f$ induced by the $tqZ$ coupling, where $q'$ and $q''$ denote the down-type quarks; $q=u,c$, and $f$ can be any possible fermions. }
\label{fig:Z_penguin}
\end{figure}

In this section, we discuss the  $tqZ$-coupling contribution to the Kaon direct CP violation. The associated  Feynman diagram is shown in Fig.~\ref{fig:Z_penguin}, where $q=u,c$; $q'$ and $q''$ are  down type quarks, and $f$ denotes any possible fermions. That is,  the involved rare $K$ and $B$ decay processes in this study are such as $K\to \pi \pi$,  $K\to \pi \nu \bar \nu$, and $K_S(B_{d}) \to \ell^+ \ell^-$. It is found that the contributions to $K_L \to \pi \ell^+ \ell^-$ and $B\to \pi \ell^+ \ell^-$ are not significant; therefore, we do not discuss the decays in this work. 

 Based on the $tqZ$ couplings shown in Eq.~(\ref{eq:L_tqZ}), the effective Hamiltonian induced by the $Z$-penguin diagram for the $K\to \pi \pi$ decays at $\mu=m_W$ can be derived as:
\begin{align}
{\cal H}_{tqZ} & = - \frac{G_F \lambda_t}{\sqrt{2}} \left( y^{Z}_3 Q_3 + y^Z_{7} Q_7 + y^Z_9 Q_9 \right) \,,
\end{align}
where $\lambda_t= V^*_{ts} V_{td}$; the operators $Q_{3,7,9}$ are the same as the SM operators and  are defined as:
 \begin{align}
 Q_{3} & = (\bar s d)_{V-A} \sum_{q'} (\bar q' q')_{V-A}\,, \nonumber \\
  Q_{7} & = \frac{3}{2} (\bar s d)_{V-A} \sum_{q'} e_{q'} (\bar q' q')_{V+A}\,, \nonumber \\
 Q_{9} & = \frac{3}{2} (\bar s d)_{V-A} \sum_{q'} e_{q'} (\bar q' q')_{V-A}\,,
 \end{align}
 with $e_{q'}$ being the electric charge of $q'$-quark, and the effective Wilson coefficients  are expressed as:
  \begin{align}
  y^Z_3 & = - \frac{\alpha}{24 \pi s^2_W} I_Z(x_t) \eta_Z \,, ~  y^Z_7  = - \frac{\alpha}{6\pi} I_Z(x_t) \eta_Z \,, \nonumber \\
  y^Z_9 & =  \left(1- \frac{1}{s^2_W}  \right)  y^Z_7 \,, ~
  \eta_Z  =  \sum_{q=u,c}  \left(   \frac{V_{qd} \zeta^{L*}_q }{V_{td}} + \frac{V^*_{qs} \zeta^L_q }{V^*_{ts}}  \right)\,, \label{eq:yZs}
  \end{align}
with $\alpha=e^2/4\pi$, $x_t=m^2_t/m^2_W$, and $s_W=\sin\theta_W$. The penguin-loop integral function is given as:
 \begin{equation}
 I_Z(x_t) = -\frac{1}{4} + \frac{x_t\, \ln x_t}{2 (x_t-1)}\approx 0.693\,.
 \end{equation}
Since $W$-boson can only couple to the left-handed quarks, the right-handed couplings $\zeta^R_{u, c}$ in the diagram have to  appear with $m_{u(c)}$ and $m_{t}$, in which the mass factors are from the mass insertion in the quark propagators inside the loop. When we drop the small factors $m_{c,u}/m_W$, the effective Hamiltonian for $K\to \pi \pi$ only depends on $\zeta^L_{u, c}$. Since $|V_{ud}/V_{td}|$ is larger than $|V_{cs}/V_{ts}|$ by   a factor of $4.67$, the dominant contribution to the $\Delta S=1$ processes is from the first term of $\eta_Z$ defined in Eq.~(\ref{eq:yZs}). In addition, $V_{ud}$ is larger than $|V_{cd}|$ by a factor of $1/\lambda \sim 4.44$; therefore, the main contribution in the first term of $\eta_Z$  comes from the $V_{ud} \zeta^{L*}_{u}/V_{td}$ effect. That is, the anomalous $tuZ$-coupling is the main effect in our study.

 Using the isospin amplitudes, the Kaon direct CP violating parameter from new physics can be estimated using~\cite{Buras:2015yba}:
 \begin{equation}
 Re\left( \frac{\epsilon'}{\epsilon}\right) \approx - \frac{  \omega }{\sqrt{2} |\epsilon_K|} \left[ \frac{Im A_0}{ Re A_0} - \frac{Im A_2}{ Re A_2}\right]\,,  \label{eq:epsilon_p}
  \end{equation}
 where $\omega = Re A_2/Re A_0 \approx 1/22.35$ denotes the $\Delta I=1/2$ rule, and $|\epsilon_K|\approx 2.228\times 10^{-3}$ is the Kaon indirect CP violating parameter. It can be seen that in addition to the  hadronic matrix element ratios,  $\epsilon'/\epsilon$ also strongly depends on the Wilson coefficients at the $\mu=m_c$ scale. It is known that the main new physics contributions to $\epsilon'/\epsilon$ are from the  $Q^{(\prime)}_6$ and $Q^{(\prime)}_8$ operators~\cite{Buras:2014sba,Buras:2015yca}. Although these operators are not generated through the $tqZ$ couplings at $\mu=m_W$ in our case, they can be induced via the QCD radiative corrections.    The Wilson coefficients at the $\mu=m_c$ scale can be obtained using the renormalization group (RG) evolution~\cite{Buchalla:1995vs}. Thus,  the induced effective Wilson coefficients for $Q_{6,8}$ operators at $\mu=m_c$ can be obtained as:
 \begin{align}
 y^Z_6(m_c) & \approx -0.08 y^Z_3 - 0.01 y^Z_7 + 0.07 y^Z_9 \,, \nonumber \\
 y^Z_8(m_c) & \approx  0.63 y^Z_7\,. 
 \end{align}
  It can be seen that $y^Z_6(m_c)$ is much smaller than $y^Z_8(m_c)$; that is, we can simply consider the  $Q_8$ operator contribution. 

 According to the $K\to \pi \pi$ matrix elements and the formulation of $Re(\epsilon'/\epsilon)$ provided in~\cite{Buras:2015yba}, the $O_8$ contribution can be written as:
 \begin{align}
 Re\left( \frac{\epsilon'}{\epsilon}\right)^Z_P & \approx - a^{(3/2)}_8  B^{(3/2)}_8 \,, \nonumber \\
 a^{(3/2)}_8 & = Im\left(\lambda_t y^Z_8(m_c)\right) \frac{r_2 \langle  Q_8 \rangle_2}{ B^{(3/2)}_8 Re A_2}\,,
 \end{align}
 where $r_2 = \omega G_F/(2 |\epsilon_K|)\approx 1.17\times 10^{-4}$ GeV$^{-2}$,  $B^{(3/2)}_8 \approx 0.76$; $Re A^{\rm exp}_{2(0)}\approx 1.21 (27.04)\times 10^{-8}$ GeV~\cite{PDG}, and the matrix element of  $\langle  Q_8 \rangle_2$ is defined as:
  \begin{equation}
  \langle  Q_8 \rangle_2 = \sqrt{2} \left( \frac{m^2_K}{m_s (\mu)+ m_d(\mu)} \right)^2 f_\pi B^{3/2}_8\,.
  \end{equation}
 Although the $Q_8$ operator can contribute to the isospin $I=0$ state of $\pi\pi$, because its effect is a factor of $15$ smaller than the isospin $I=2$ state, we thus neglect its contribution. 

\begin{figure}[phtb]
\includegraphics[scale=0.7]{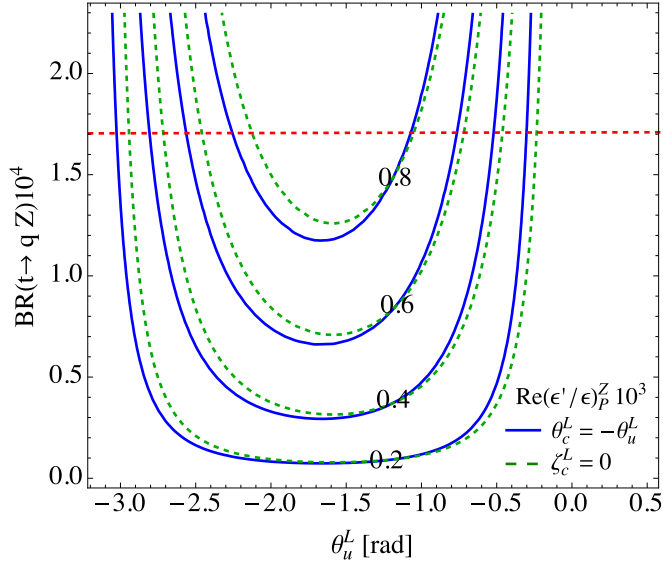}
 \caption{ Contours for $Re(\epsilon'/\epsilon)^Z_P$ (in units of $10^{-3}$) as a function of  $BR(t\to q Z)$ and $\theta^L_u$, where the solid and dashed lines denote the  $\theta^L_c=-\theta^L_u$ and $\zeta^L_c=0$ results, respectively. The $BR(t\to q Z)$ is defined as the minimal one between $BR(t\to u Z)$ and $BR(t\to c Z)$. The horizontal dashed line (red) is the current upper limit of $BR(t\to q Z)$. }
\label{fig:ep_e}
\end{figure}

Since the  $t\to (u, c)Z$ decays  have not yet been observed, in order to simplify their correlation to $\epsilon'/\epsilon$, we use $BR(t\to q Z)\equiv {\rm Min}(BR(t\to cZ), \, BR(t\to u Z))$ instead of $BR(t\to u(c) Z)$ as the upper limit. The contours for $Re(\epsilon'/\epsilon)^Z_P$ ( in units of $10^{-3}$) as a function of $BR(t\to q Z)$ and $\theta^L_u$ are shown in Fig.~\ref{fig:ep_e}, where the solid and dashed lines denote the results with  $\theta^L_c = -\theta^L_u$ and $\zeta^L_c=0$, respectively, and the horizontal dashed line is the current upper limit of $BR(t\to q Z)$. It can be seen that the Kaon direct CP violation arisen from the anomalous $t u Z$-coupling  can reach $0.8 \times 10^{-3}$, and the contribution from $tcZ$-coupling is only a minor effect. When the limit of $t\to qZ$ approaches $BR(t\to qZ)\sim 0.5 \times 10^{-4}$, the induced $\epsilon'/\epsilon$ can be as large as $Re(\epsilon'/\epsilon)^Z_P \sim 0.4 \times 10^{-3}$.

\section{Z-penguin induced (semi)-leptonic $K$ and $B$ decays and numerical analysis}

 The same Feynman diagram as that in Fig.~\ref{fig:Z_penguin} can be also applied to the rare leptonic and semi-leptonic $K(B)$ decays when $f$ is a neutrino or a charged lepton. Because $|V_{us}/V_{ts}| \ll |V_{cs}/V_{ts}| \sim |V_{us}/V_{td}| \ll |V_{ud}/V_{td}|$, it can be  found that the anomalous $tu(c)Z$-coupling contributions  to the $b\to s \ell \bar \ell$ ($\ell= \nu, \ell^-)$ processes  can deviate from the SM result being less than $7\%$ in terms of  amplitude. However, the influence of the $tuZ$ coupling on $d \to s \ell \bar\ell$ and $b\to d \ell \bar \ell$ can be over $20\%$ at the amplitude level.  Accordingly, in the following analysis, we concentrate  the study on the rare decays, such as $K\to \pi \nu\bar\nu$, $K_{S}\to \mu^+ \mu^-$, and $B_d \to \mu^+ \mu^-$, in which the channels are sensitive to the new physics effects and are theoretically clean.  
 
 According to the formulations in~\cite{Bobeth:2017ecx}, we write the effective Hamiltonian for $d_i \to d_j \ell \bar \ell$ induced by the $tuZ$ coupling as:
\begin{align}
{\cal H}_{d_i \to d_j \ell \bar \ell} &=-\frac{G_F V^*_{td_j } V_{td_i} }{\sqrt{2} } \frac{\alpha}{\pi }  C^Z_L  [ \bar d_j  \gamma_\mu P_L d_i] [\bar\nu \gamma^\mu (1-\gamma_5)\nu ]  \nonumber \\
& - \frac{G_F V^*_{td_j } V_{td_i} }{\sqrt{2}} \frac{\alpha}{ \pi } \bar d_j \gamma_\mu P_L d_i \left[ C^Z_9  \bar \ell \gamma^\mu \ell + C^Z_{10}  \bar \ell \gamma^\mu \gamma_5 \ell \right]\,,   \label{eq:Heff}
\end{align}
where we have ignored the small contributions from the $tcZ$-coupling; $d_i \to d_j$ could be the $s\to d$ or $b\to d$ transition, and the effective Wilson coefficients are given as:
 \begin{align}
 C^Z_L =   C^Z_{10} & \approx  \frac{I_Z(x_t) \zeta^{L}_u}{4s^2_W} \frac{V^*_{ud}}{V^*_{td}} \,, ~ C^Z_{9} \approx  C^Z_L \left( -1 + 4 s^2_W \right)\,.
 \end{align}
Because $-1+4s^2_W \approx -0.08$, the $C^Z_9$ effect  can indeed be neglected.

 Based on the interactions in Eq.~(\ref{eq:Heff}), the BRs for the $K_L \to \pi^0 \nu \bar\nu$ and $K^+\to \pi^+ \nu \bar\nu$ decays can be formulated as~\cite{Buras:2015yca}:
\begin{align}
BR(K_L \to \pi^0 \nu \bar \nu) & = \kappa_L \left| \frac{Im\, X_{\rm eff} }{\lambda^5} \right|^2 \,, \nonumber \\
BR(K^+ \to \pi^+ \nu \bar\nu) & = \kappa_+ (1+\Delta _{\rm EM}) \left[ \left| \frac{Im\, X_{\rm eff}}{\lambda^5 } \right|^2 +\left| \frac{Re\, \lambda_c}{\lambda} P_c(X) + \frac{Re\, X_{\rm eff}}{\lambda^5}\right|^2 \right]\,,
\end{align}
where $\lambda_c= V^*_{cs} V_{cd}$, $\Delta_{EM}=-0.003$; $P_c(X)=0.404\pm 0.024$ denotes the charm-quark contribution~\cite{Isidori:2005xm,Mescia:2007kn}; the values of $\kappa_{L}$ and $\kappa_+$ are respectively given as $\kappa_L=(2.231 \pm 0.013)\times 10^{-10}$ and $\kappa_{+}=(5.173\pm 0.025)\times 10^{-11}$\,, and  $X_{\rm eff}$ is defined as:
 \begin{equation}
 X_{\rm eff} = \lambda_t \left( X^{\rm SM}_L -  s^2_W C^{Z*}_L \right)\,,
 \end{equation}
with $ X^{\rm SM}_L =1.481 \pm 0.009$~\cite{Buras:2015yca}.  Since $K_L\to \pi^0 \nu \bar\nu$ is a CP violating process, its BR only depends on the imaginary part of $X_{\rm eff}$.  Another important CP violating process in $K$ decay is $K_S\to \mu^+ \mu^-$, where its BR from the SD contribution can be expressed as~\cite{Bobeth:2017ecx}:
\begin{align}
BR(K_S\to \mu^+ \mu^-)_{\rm SD} & = \tau_{K_S} \frac{G^2_F \alpha^2}{8\pi^3} m_K f^2_K m^2_\mu \sqrt{1- \frac{4m^2_\mu}{m^2_{K}}}
 \left| Im [\lambda_t \left( C^{\rm SM}_{10} + C^{Z*}_{10} \right)]\right|^2\,, 
\end{align}
with $C^{\rm SM}_{10}\approx -4.21$. Including the LD effect~\cite{Ecker:1991ru,Isidori:2003ts}, the BR for $K_S \to \mu^+ \mu^-$ can be estimated using $BR(K_S\to \mu^+ \mu^-)_{\rm LD+SD}\approx 4.99_{\rm LD} \times 10^{-12}+ BR(K_S\to \mu^+ \mu^-)_{\rm SD}$~\cite{DAmbrosio:2017klp}. Moreover, it is found that the effective interactions in Eq.~(\ref{eq:Heff}) can  significantly affect the $B_d \to \mu^+ \mu^-$ decay, where its BR can be derived as:
\begin{align}
 BR(B_d \to \mu^+ \mu^-) & = \tau_{B} \frac{G^2_F \alpha^2}{16\pi^3} m_{B} f^2_B m^2_\mu \left(1-  \frac{2 m^2_\ell}{m^2_B} \right)\sqrt{1- \frac{4 m^2_\mu}{m^2_{B}}} \nonumber \\
 & \times \left| V^*_{td} V_{tb} \left( C^{\rm SM}_{10} + C^Z_{10} \right)\right|^2\,. 
 \end{align}
Because $B_d \to \mu^+ \mu^-$ is not a pure CP violating process, the BR involves both the real and imaginary part of $V^*_{td} V_{tb} \left( C^{\rm SM}_{10} + C^Z_{10} \right)$. Note that the associated Wilson coefficient in $B_d \to \mu^+ \mu^-$ is $C^{Z}_{10}$, whereas  it is $C^{Z*}_{10}$ in the $K$ decays. 

\begin{figure}[phtb]
\includegraphics[scale=0.5]{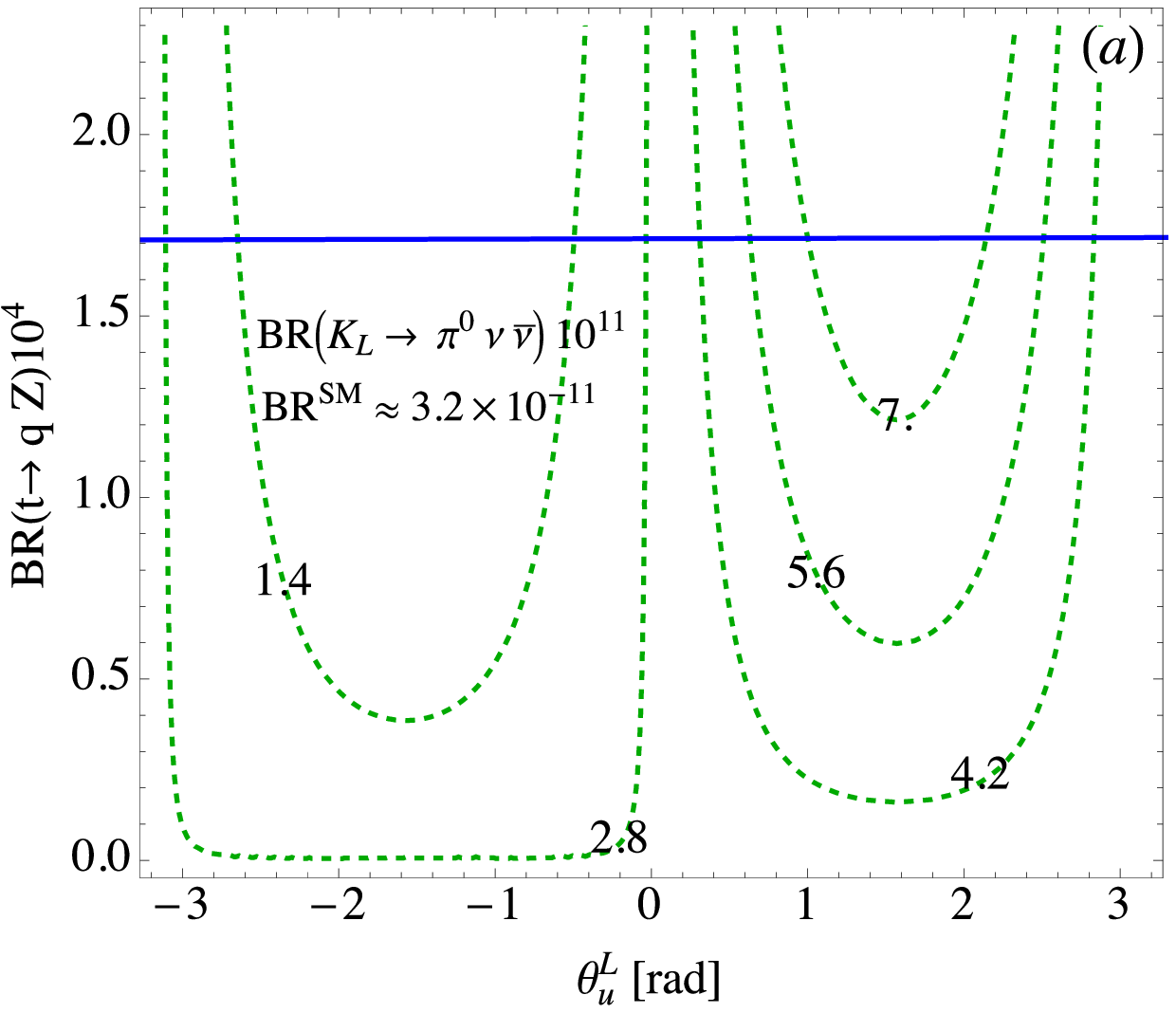}
\includegraphics[scale=0.5]{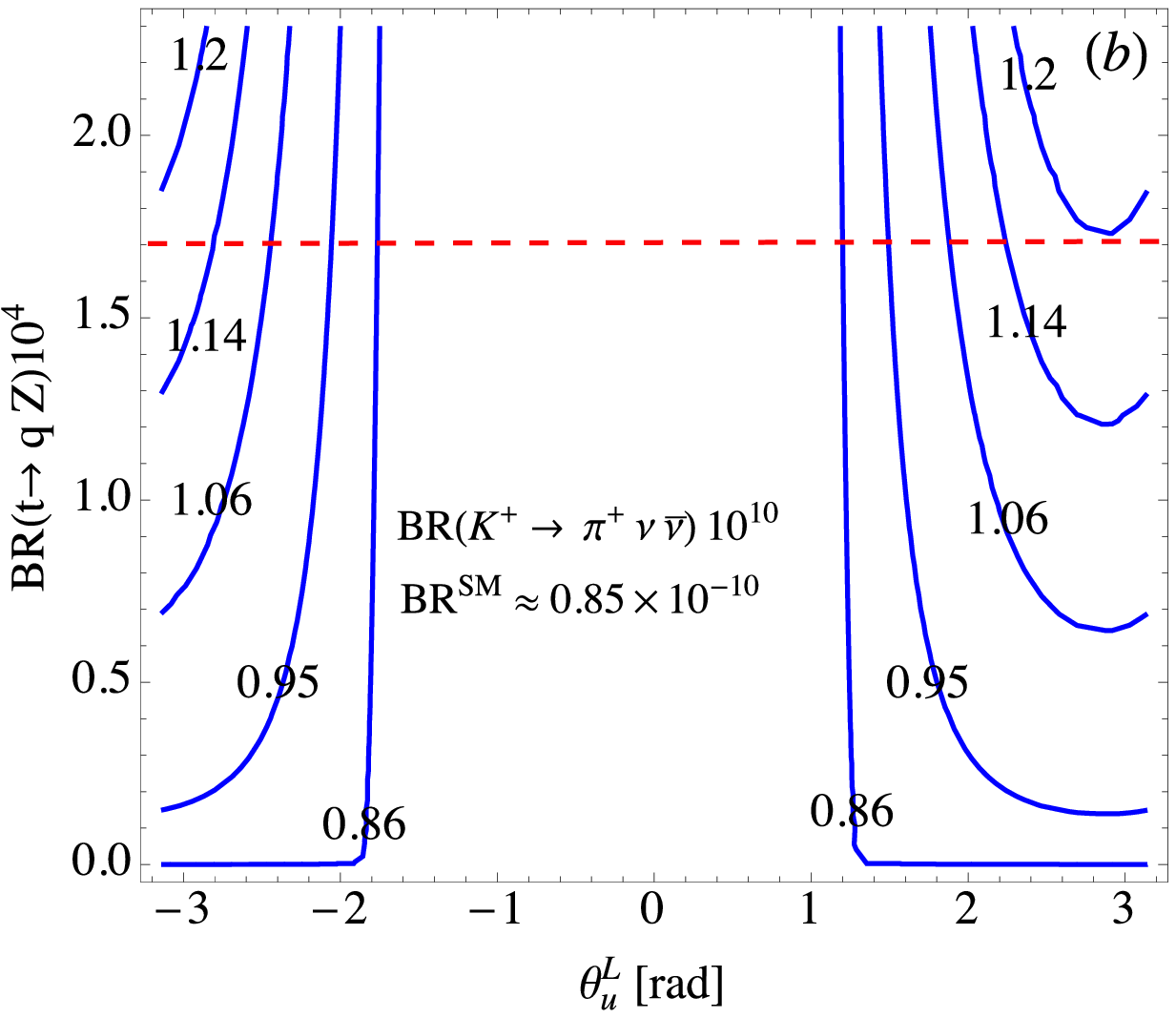}
\includegraphics[scale=0.5]{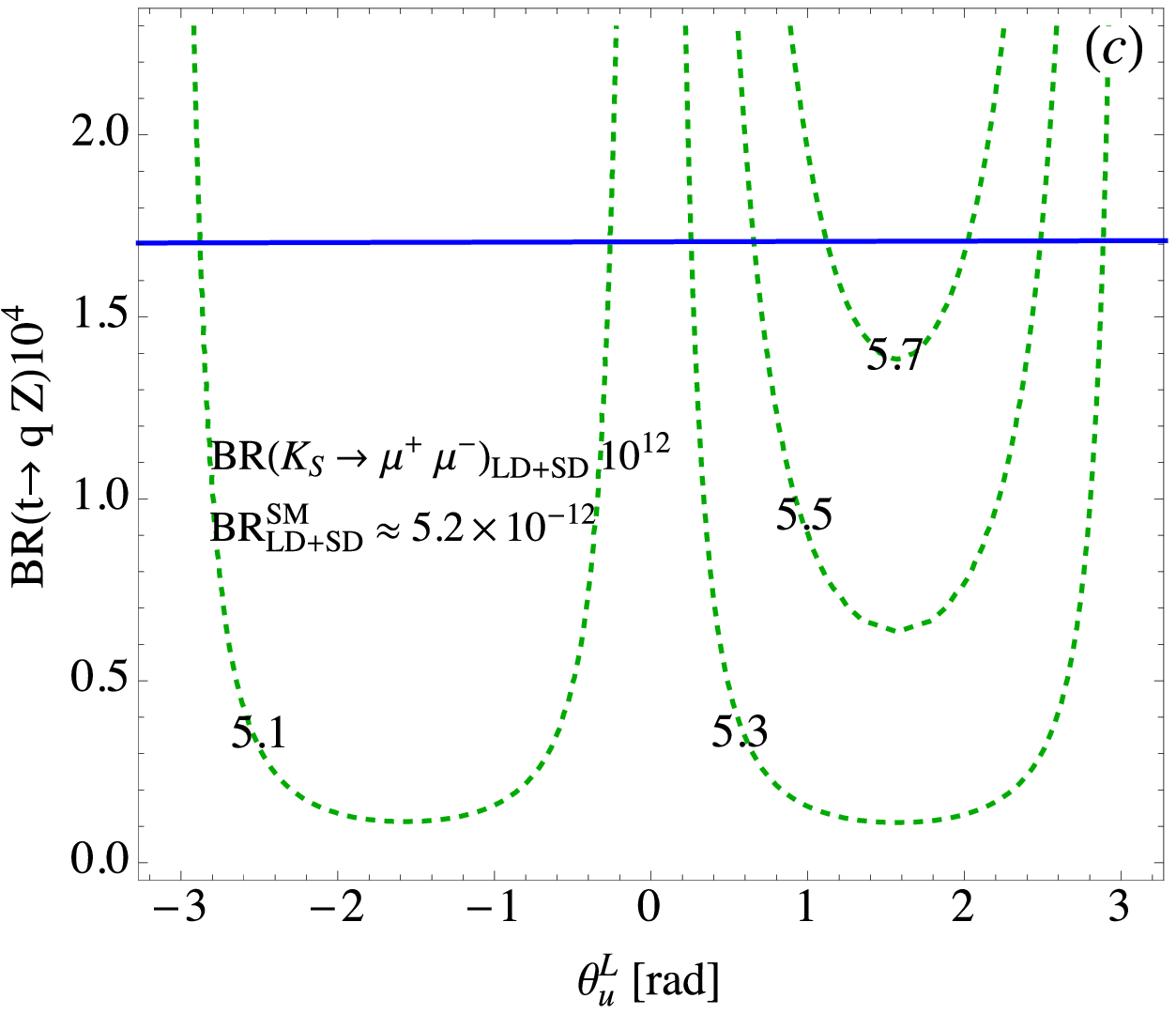}
\includegraphics[scale=0.5]{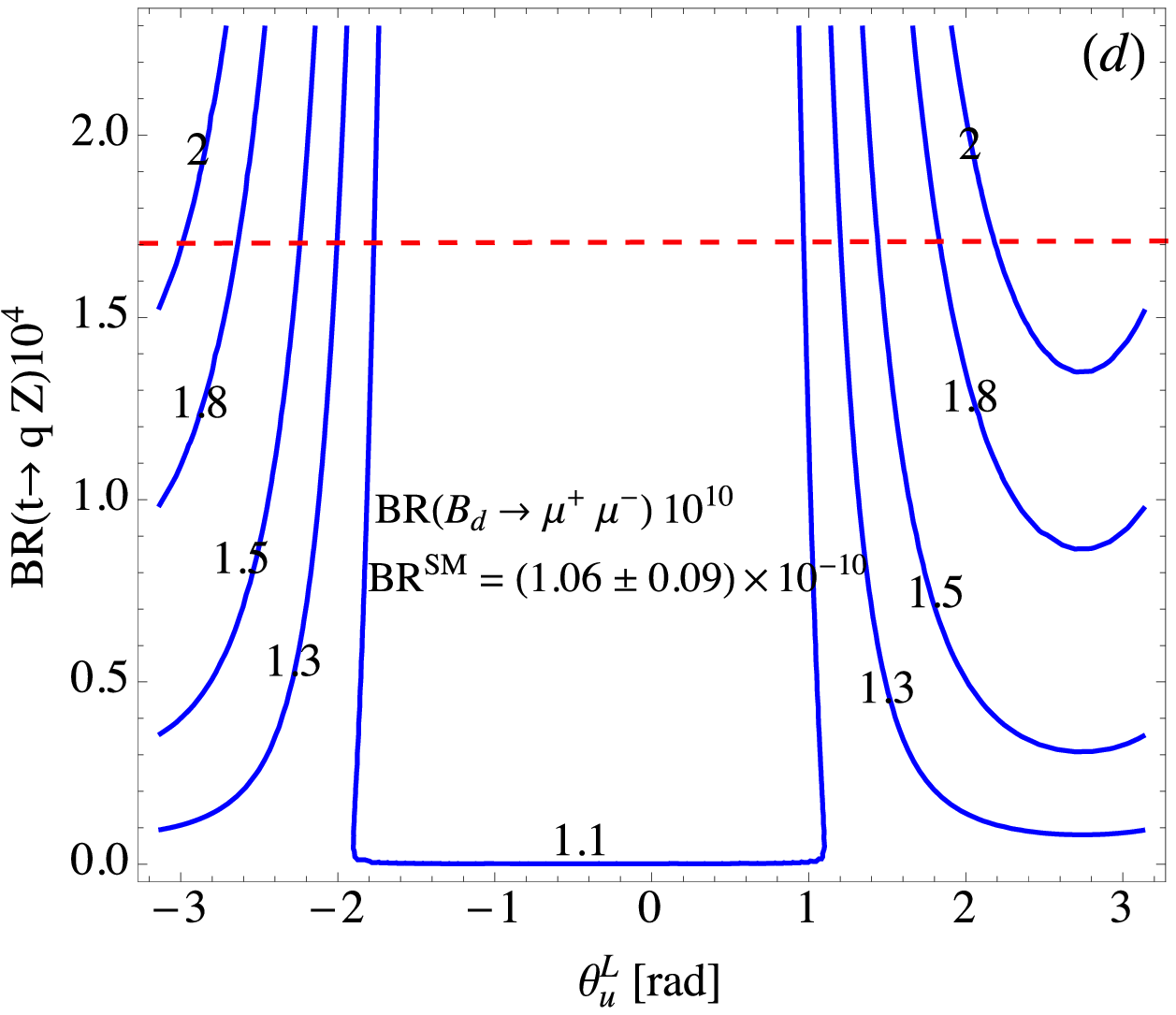}
 \caption{  Contours of the branching ratio as a function of $BR(t\to u Z)$ and $\theta^L_u$ for (a) $K_L\to \pi^0 \nu \bar\nu$, (b) $K^+ \to \pi^+ \nu \bar\nu $, (c) $K_S\to \mu^+ \mu^{-}$, and (d) $B_d \to \mu^+ \mu^{-}$, where the corresponding SM result is also shown in each plot. The long-distance effect has been included in the $K_S\to \mu^+ \mu^-$ decay.  }
\label{fig:K_B}
\end{figure}

 After formulating the BRs for the investigated processes, we now numerically analyze the  $tuZ$-coupling effect on these decays. Since the involved parameter is the complex  $\zeta^L_u=|\zeta^L_u| e^{-i\theta^L_u}$,  we take $BR(t\to uZ)$ instead of $|\zeta^L_u|$. Thus, we show  $BR(K_L \to \pi^0 \nu \bar\nu)$ (in units of $10^{-11})$ as a function of $BR(t\to u Z)$ and $\theta^L_u$ in Fig.~\ref{fig:K_B}(a), where the CP phase is taken in the range of  $\theta^L_u=[-\pi, \pi]$;  the SM result is  shown in the plot, and the horizontal line denotes the current upper limit of $BR(t\to u Z)$. It can be clearly seen that $BR(K_L \to \pi^0 \nu \bar\nu)$ can be enhanced to $7\times 10^{-11}$    in  $\theta^L_u >0$ when  $BR(t\to uZ)<1.7 \times 10^{-4}$ is satisfied. Moreover,   the result of $BR(K_L\to \pi^0 \nu \bar\nu)\approx 5.3 \times 10^{-11}$ can be achieved when $BR(t\to u Z)=0.5\times 10^{-4}$ and $\theta^u_L\ =2.1$ are used. Similarly,  the influence of  $\zeta^L_u$ on $BR(K^+\to \pi^+ \nu \bar\nu)$ is shown in Fig.~\ref{fig:K_B}(b). 
Since $BR(K^+\to \pi^+ \nu \bar\nu)$ involves the real and imaginary parts of $X_{\rm eff}$, unlike the $K_L \to \pi^0 \nu \bar\nu$ decay, its BR cannot be enhanced manyfold due to the dominance of the real part. Nevertheless, the BR of $K^+\to \pi^+ \nu \bar\nu$ can  be maximally enhanced by  $38\%$; even, with $BR(t\to u Z)=0.5\times 10^{-4}$ and $\theta^u_L= 2.1$,  the $BR(K^+\to \pi^+ \nu \bar\nu)$  can still exhibit an increase of  $15\%$. It can be also found that in addition to $|\zeta^L_u|$, the BRs of $K\to \pi \nu \bar\nu$ are also sensitive to the $\theta^L_u$ CP-phase.  Although the observed $BR(K\to \pi \nu \bar \nu)$ cannot constrain $BR(t\to u Z)$, the allowed  range of $\theta^L_u$ can be further limited. 

For the $K_S \to \mu^+ \mu^-$ decay, in addition to the SD effect,  the LD effect, which arises from the absorptive part of $K_S \to \gamma \gamma \to \mu^+ \mu^-$,  predominantly  contributes to the $BR(K_S \to \mu^+ \mu^-)$. Thus, if the new physics contribution is much smaller than the LD effect, the  influence on  $BR(K_S\to \mu^+ \mu^-)_{\rm LD +SD}=BR(K_S\to \mu^+ \mu^-)_{\rm LD }+ BR(K_S\to \mu^+ \mu^-)_{\rm SD}$ from new physics may not be so significant. In order to show the $tuZ$-coupling effect, we plot the contours for $BR(K_S\to \mu^+ \mu^-)_{\rm LD +SD}$ ( in units of $10^{-12}$)  in Fig.~\ref{fig:K_B}(c). From the result, it can be clearly seen that $BR(K_S\to \mu^+ \mu^-)_{\rm LD +SD}$ can be at most enhanced by $11\%$ with respect to the SM result, whereas the BR can  be enhanced only  $\sim 4.3\%$ when $BR(t\to uZ)=0.5\times 10^{-4}$ and $\theta^L_u=2.1$ are used.  We note that the same new physics effect also contributes to $K_L\to \mu^+ \mu^-$. Since the SD contribution to $K_L\to \mu^+ \mu^-$ is smaller than the SM SD effect by one order of magnitude, we skip to show the case for the $K_L \to \mu^+ \mu^-$ decay. 

As discussed earlier that the $tcZ$-coupling contribution to the $B_s \to \mu^+ \mu^-$ process is small; however,  similar to the case in $K^+ \to \pi^+ \nu \bar\nu$ decay, the BR of $B_d \to \mu^+ \mu^-$  can be significantly enhanced through the anomalous $tuZ$-coupling. We show the contours of $BR(B_d \to \mu^+ \mu^-)$ ( in units of $10^{-10})$ as a function of $BR(t\to uZ)$ and $\theta^L_u$  in Fig.~\ref{fig:K_B}(d). It can be seen that the maximum of the allowed $BR(B_d \to \mu^+ \mu^-)$ can reach $1.97 \times 10^{-10}$, which is a factor of 1.8 larger than the SM result of $BR(B_d \to \mu^+ \mu^-)^{\rm SM}\approx 1.06 \times 10^{-10}$. Using $BR(t\to u Z)=0.5\times 10^{-4}$ and $\theta^L_u=2.1$, the enhancement factor to $BR(B_d \to \mu^+ \mu^-)^{\rm SM}$ becomes $1.38$.  Since the maximum of $BR(B_d \to \mu^+ \mu^-)$ has been close to the ATLAS upper bound of $2.1\times 10^{-10}$, the constraint from the rare $B$ decay measured in the LHC could further constrain the allowed range  of  $\theta^L_u$

\section{Summary}

We studied the impacts of the anomalous $tqZ$ couplings  in the low energy physics, especially the $tuZ$ coupling. It was found that  the anomalous coupling can have significant contributions to $\epsilon'/\epsilon$, $BR(K\to \pi \nu \bar\nu)$, $K_S \to \mu^+ \mu^-$, and $B_d \to \mu^+ \mu^-$. Although these decays have not yet been observed in experiments, with the exception of $\epsilon'/\epsilon$, their designed experiment sensitivities are good enough to test the SM. It was found that using the sensitivity of $BR(t\to uZ)\sim 5\times 10^{-5}$ designed in HL-LHC, the resulted $BR(K\to \pi \nu \bar\nu)$ and $BR(B_d \to \mu^+ \mu^-)$ can be examined by the NA62, KOTO, KELVER, and LHC experiments. 

According to our study,  it was found that we cannot simultaneously enhance $Re(\epsilon'/\epsilon)$, $BR(K_L \to \pi^0 \nu \bar\nu)$, and $BR(K_S \to \mu^+ \mu^-)$ in the same region of the CP violating phase, where the positive $Re(\epsilon'/\epsilon)$ requires $\theta^L_u < 0$, but the large $BR(K_L \to \pi^0 \nu \bar\nu)$ and $BR(K_S \to \mu^+ \mu^-)$ have to rely on $\theta^L_u > 0$. Since $BR(K^+ \to \pi^+ \nu \bar \nu)$ and $BR(B_d \to \mu^+ \mu^-)$ involve both real and imaginary parts of Wilson coefficients, their BRs are not sensitive to the sign of $\theta^L_u$. Hence,  $Re(\epsilon'/\epsilon)$, $BR(K^+ \to \pi^+ \nu \bar \nu)$ and $BR(B_d \to \mu^+ \mu^-)$ can be enhanced at the same time.  


\appendix

\section{Anomalous gauge couplings from the SM-EFT}

{

If we take the SM as an effective theory at the electroweak scale, the new physics effects should appear in terms of higher dimensional operators when the heavy fields above electroweak scale are integrated out. Thus, the effective Lagrangian with respect to the SM gauge symmetry can be generally expressed as~\cite{Buchmuller:1985jz,AguilarSaavedra:2009mx,Grzadkowski:2010es}:
 \begin{equation}
 {\cal L}= {\cal L}^{(4)}_{\rm SM} + \frac{1}{\Lambda} \sum_k C^{(5)}_k Q^{(5)}_k + \frac{1}{\Lambda^2} \sum_k C^{(6)}_k Q^{(6)}_k + ...\,,
 \end{equation}
where ${\cal L}^{(4)}_{\rm SM}$ is the original SM; $Q^{(n)}_k$ are the dimension-$n$ effective operators, and $C^{(n)}_k$ are the associated Wilson coefficients. The top flavor-changing anomalous couplings can be generated from the dimension-6 operators, where based on the notations in~\cite{Grzadkowski:2010es},  the relevant operators in our study can be written as~\cite{Grzadkowski:2010es}:
 \begin{align}
 {\cal L} & \supset \frac{1}{\Lambda^2} \left\{
   \left(\vpj \right) \left( Q_L \gamma^\mu C^{(1)}_{\phi q} Q_L\right) + \left( \vpjt \right) \left( Q_L \tau^{I}\gamma^\mu C^{(3)}_{\phi q} Q_L \right) \right.\nonumber \\
&   \left. +  \left(\vpj \right) \left[ U_R \gamma^\mu C_{\phi u} U_R + D_R \gamma^\mu C_{\phi d} D_R \right] + \left(\tvp^\dag iD_\mu \vp \right) \left(U_R \gamma^\mu C_{\phi u d} D_R \right)
 \right\}\,, \label{eq:SM_EFT}
 \end{align}
 where $\vp$ denotes the SM Higgs doublet, $Q^T_L=( U_L , D_L)$ is left-handed quark doublet, $D_\mu \varphi$ is the covariant derivative acting on $\varphi$, $\tau^I$ are the Pauli matrices; $\tvp = i\tau_2 \vp^*$, $\vp^\dag\Db_\mu \vp = (D_\mu \vp)^\dag \vp$, and 
 \begin{align}
 \vpj  = i \vp^\dag \left( D_\mu - \Db_\mu \right) \vp\,, ~ \vpjt = i \vp^\dag \left( \tau^I D_\mu - \Db_\mu \tau^I \right) \vp\,.
 \end{align}
The flavor indices are suppressed; therefore, the Wilson coefficients $\{ C_{i}\}$ are $3\times 3$ matrices.  Since the top anomalous gauge couplings in this study are mainly related to the left-handed couplings, in the following discussions, we focus  on the couplings to the left-handed quarks. 

After electroweak symmetry breaking, the relevant  $Z$ and $W$ gauge couplings to the quark weak eigenstates in Eq.~(\ref{eq:SM_EFT}) can be formulated as:
 \begin{align}
 {\cal L} & \supset \frac{g }{2c_W} \frac{v^2}{\Lambda^2}  \left[ \bar d_L \gamma^\mu \left(C^{(1)}_{\phi q}+C^{(3)}_{\phi q} \right) d_L  +  \bar u_L \gamma^\mu \left(C^{(1)}_{\phi q}-C^{(3)}_{\phi q} \right) u_L \right] Z_\mu \nonumber \\
 & - \frac{g }{\sqrt{2}} \frac{v^2}{\Lambda^2}  \left[  \bar u_L \gamma^\mu C^{(3)}_{\phi q} d_{L} W^+_\mu + \bar d_L \gamma^\mu C^{(3)}_{\phi q} u_{L} W^-_\mu   \right]  + H.c,\label{eq:red_EFT}
 \end{align}
 where $<\varphi>=v/\sqrt{2}$ is the vacuum expectation value (VEV) of $\varphi$.
It can be seen that the $Z$ couplings to the down-type quarks can be removed if we assume $C^{(1)}_{\phi q}= - C^{(3)}_{\phi q} \equiv - C_{qL}$. Under such circumstance, the FCNCs at the tree level could only occur in the up-type quarks. In order to use the physical quark states to express Eq.~(\ref{eq:red_EFT}), we introduce the unitary matrices $U^{u,d}_{L, R}$ to diagonalize the quark mass matrices. Thus, defining $C'_{qL} = V^u_L C_{qL} V^{u\dagger}_L$, Eq.~(\ref{eq:red_EFT}) can be written as:
 \begin{align}
  {\cal L} & \supset - \frac{g }{2c_W} \frac{v^2}{\Lambda^2}    \bar u_L \gamma^\mu \xi_{qL} u_L\,  Z_\mu  - \frac{g }{2 \sqrt{2}} \frac{v^2}{\Lambda^2}   \bar u_L \gamma^\mu \xi_{qL} V_{\rm CKM} d_{L}  \,   W^+_\mu   + H.c.\,,\label{eq:WZ_EFT}
 \end{align}
where $\xi_{qL} = 2 C'_{qL} + 2 C'^{\dagger}_{qL}$, and $V=V^u_L V^{d\dagger}_L$ is the Cabibbo-Kobayashi-Maskawa (CKM) matrix. It can be seen that the anomalous gauge couplings in the neutral current interactions are strongly correlated with those in the charged-current interactions.

It is known that the CKM matrix has a hierarchical structure, such as  $V_{11(22,33)} \sim 1 $, $|V_{12(21)}| \sim \lambda$, $|V_{23(32)}| \sim \lambda^2$, and $|(V_{13(31)}| \sim \lambda^3$, where $\lambda\approx 0.22$ is the Wolfenstein parameter~\cite{Wolfenstein:1983yz}.  Since each CKM matrix element is measured well, it is necessary to examine if the sizable  $t\to q Z$ FCNCs are excluded by the experimental measurements, which are dictated by  the charged current interactions. Thus, in the following analysis, we concentrate on the modifications of $V_{ub}$, $V_{ts}$, and $V_{td}$. First, we consider $(\xi_{qL} V)_{ub}$ for the $b\to u$ transition effect and decompose it as:
 \begin{align}
 (\xi_{qL} V)_{ub} & = (\xi_{qL})_{uu} V_{ub} + (\xi_{qL})_{uc} V_{cb} + (\xi_{qL})_{ut} V_{tb} \nonumber \\
  & \approx (\xi_{qL})_{ut} V_{tb}\,,
 \end{align}
where  the $V_{ub, cb}$ terms in the second line are dropped due to $V_{ub,cb}\ll V_{tb}$. In order to obtain a small effect in the $b\to u$ transition, we have to require $v^2(\xi_{qL})_{ut}/(\Lambda^2)$ to be much less than  $0.02$, which is the current upper limit shown in Eq.~(\ref{eq:zeta_upper_limit}). Similarly, the $(\xi_{qL} V_{\rm CKM})_{ts(td)}$ factors  can be expressed in terms of $\lambda$ as:
 \begin{align}
 (\xi_{qL} V)_{ts} &  \approx  \lambda (\xi_{qL})_{tu} + (\xi_{qL})_{tc} - \lambda^2  (\xi_{qL})_{tt} \,, \nonumber \\
  (\xi_{qL} V)_{td} &  \approx    (\xi_{qL})_{tu} - \lambda (\xi_{qL})_{tc} + \lambda^3 (\xi_{qL})_{tt}\,.
 \end{align}
 If we take  $(\xi_{qL})_{tu} \sim \lambda (\xi_{qL})_{tc} - \lambda^3 (\xi_{qL})_{tt}$, i.e., the $(\xi_{qL} V)_{td}$ effect  is suppressed,  $(\xi_{qL} V)_{ts}$ can be rewritten as:
  \begin{align}
   (\xi_{qL} V)_{ts} \approx  (1+ \lambda^2) \left[ (\xi_{qL})_{tc} - \lambda^2  (\xi_{qL})_{tt} \right] \sim \frac{(\xi_{qL} )_{tu}}{\lambda}\,.
  \end{align}
 Because $(\xi_{qL})_{tu,tc,tt}$ are taken as the  free parameters, we have the degrees of freedom to obtain $ v^2 |(\xi_{qL} V)_{ts} |/(2\Lambda^2)< |V_{ts}| \sim \lambda^2$ without $(\xi_{qL})_{tu(tc)} \ll1$. Using the result, we can obtain $|\zeta^L_{u}|=v^2 |(\xi_{qL} )_{tu}|/\Lambda^2 < 0.021$, where the upper limit is consistent with that shown in Eq.~(\ref{eq:zeta_upper_limit}).  Hence, although  $v^2 (\xi_{qL})_{ut}/\Lambda^2$ in a general SM-EFT is bounded by the measured CKM matrix elements, $\zeta^L_u =v^2 (\xi_{qL})_{tu}/\Lambda^2$ could be a free parameter and  $\zeta^L_u <0.021$ is still allowed. 
 
}

\section*{Acknowledgments}

This work was partially supported by the Ministry of Science and Technology of Taiwan,  
under grants MOST-106-2112-M-006-010-MY2 (CHC).

\end{document}